\documentclass[twocolumn,aps,preprintnumbers,amsmath,amssymb,floatfix]{revtex4}
\usepackage{graphicx}

\begin{document}

\title{Properties of YBa$_2$Cu$_3$O$_{7-\delta}$ films grown by pulsed laser deposition on CeO$_2$-buffered sapphire}
\author{I.~Abaloszewa$^1$, P.~Gier\l{}owski$^1$, A.~Abaloszew$^1$, I.~Zaytseva$^1$, M.~Aleszkiewicz$^1$, Y.~Syryanyy$^1$\footnote{Present address: National Centre for Nuclear Research, A. So\l{}tana 7, 05-400, Otwock-\'{S}wierk, Poland.}, V.~Bezusyy$^1$, A.~Malinowski$^1$, M.Z.~Cieplak$^1$, M.~Jaworski$^1$, M.~Konczykowski$^{2}$, A.~Abramowicz$^3$, {\v S}.~Chromik$^4$, E.~Dobro{\v c}ka$^4$}
\affiliation{$^1$Institute of Physics, Polish Academy of Sciences, 02-668 Warszawa, Al. Lotnik{\'o}w 32/46, Poland}
\affiliation{$^2$Laboratoire des Solides Irradies CEA/DSM/IRAMIS \& CNRS UMR7642,  Ecole Polytechnique,  91128 Palaiseau, France}
\affiliation{$^3$Warsaw University of Technology, Institute of Electronic Systems, ul. Nowowiejska 15/19,
00-665 Warszawa, Poland}
\affiliation{$^4$Institute of Electrical Engineering, Slovak Academy of Sciences, Dúbravská cesta 9, 84104 Bratislava, Slovak Republic}
\date{\today}

\begin{abstract}

In the present work we study the growth by pulsed laser deposition of YBa$_2$Cu$_3$O$_{7-\delta}$ (YBCO) films on the r-cut sapphire substrates. To improve the matching of the lattice parameters between the substrate and the film we use CeO$_{2}$ buffer layer, recrystallized prior to the deposition of YBCO. The optimal thickness and temperature of recrystallization of the buffer layer is first determined using atomic force microscopy (AFM) and X-ray diffraction. Next, we use the AFM to examine the dependence of YBCO film roughness on the film thickness, and we study the homogeneity of magnetic flux penetration into the films by magneto-optical imaging. We find that the superconducting critical temperature and critical current density of these films are very similar to those of YBCO films grown on well-matched substrates. It appears that the microstructure of YBCO films is affected by structural defects in the buffer layer as well as variations in oxygen deficiency, which results in high values of critical current density suitable for application.

\end{abstract}
\maketitle

\section{Introduction}

Intensive development of superconducting microwave filters based on high-temperature superconducting films has taken place in recent years, with main attention focused on YBCO films, which display very low surface resistance in comparison to the best conducting normal metals \cite{Yamasaki, Zhao, SIMON, Lorenz}. One of the best substrates for films used in such applications is a CeO$_2$-buffered sapphire, which exhibits low microwave losses in devices \cite{Wu, Chromik1,Ohki}.  The CeO$_2$ buffer layer  prevents chemical reaction and improves matching of the lattice parameters between the substrate and the superconductor layer. The cerium dioxide has fluorite structure with a cubic lattice constant $a_{\rm{buf}} = 5.411$ \AA{} \cite{Wu}. Epitaxial YBCO film grows on CeO$_2$ buffered r-cut sapphire rotated 45$^\circ$ in the buffer basal plane and has small lattice mismatch with the $a_{\rm{buf}}$ parameter, equal to $0.16\%$ and $1.7\%$, along the \emph{a} and \emph{b} axes of the YBCO, respectively \cite{Wu}. In addition, the CeO$_2$ is cheap, chemically stable, and displays both high thermal conductivity and high melting temperature (2400 $^\circ$C). However, a wide variation in the properties of YBCO films deposited on CeO$_2$ buffer layers is observed, depending on the growth technique. The thermal expansion coefficient of sapphire is about two times smaller than that of YBCO \cite{Nie}, what often causes cracking of the film during cooling after deposition \cite{microcracking, Kastner}. While the effect of YBCO film thickness and growth parameters on the occurrence of microcracks has been reported previously \cite{Develos-Bagarinao}, the parameters of CeO$_2$ buffer layer, optimal for avoiding the cracking, have not been discussed in detail. Earlier we reported briefly on the results of our research on the growth of YBCO thin films on CeO$_2$ buffered sapphire substrates by pulsed laser deposition (PLD) method \cite{abali, abali1}, and in the present paper we provide a systematic summary description of these studies. We find the optimal thickness and parameters of buffer growth and recrystallization and define the optimal conditions for YBCO film growth to prevent film cracking and to maximize superconducting state parameters, namely, the superconducting critical temperature $T_{\rm{c0}}$ and the critical current density $j_{\rm{c}}$. The reproducibility of the film growth process that results in samples with high values of $T_{\rm{c0}}$ and $j_{\rm{c}}$ is found to be about 70 \% of the overall number of deposited films. We discuss the possible origins of high values of $j_{\rm{c}}$, which is most likely caused by strong vortex pinning by correlated defects. Finally, we show that such films deposited on substrates of larger size can be successfully used to build filters for microwave applications.

\section{Film preparation and measurement details}

The cerium dioxide layers were prepared as follows. First, the r-cut $(0 \bar{2} 1 1)$ sapphire substrates were annealed at 1000 $^\circ$C in the air for 30 minutes in order to smooth their surface before CeO$_2$ deposition. The CeO$_2$ thin films were grown on the $5 \times 5$ mm$^2$ substrates by PLD from ceramic target mounted on a rotating holder opposite to a substrate. The PLD was performed by Quanta-Ray Pro 350-10 (Spectra Physics, USA) Nd:YAG laser with the 4th harmonic generation (266 nm wavelength), using a pulse duration of 9 ns, repetition rate of 1 Hz and energy density of 1.5 J/cm$^2$ on the target surface. During the deposition, the substrate temperature was maintained at 785 $^\circ$C and the oxygen pressure in the chamber was 400 mTorr. After cooling to room temperature at a rate of 20 K/min and oxygen pressure of 300 Torr, the substrates with buffer layers of CeO$_2$ were removed from the deposition chamber and CeO$_2$ was recrystallized inside the tube furnace at high temperatures (800-1200 $^\circ$C) in the oxygen flow for one hour.

After recrystallization, the YBCO films were deposited on the top of CeO$_2$ layers using identical PLD parameters as for the buffer layers. We prepared CeO$_2$ films with thicknesses in the range of 30-90 nm and YBCO films in the range of 100-300 nm. We used our custom-made polycrystalline YBCO target characterized by excellent superconducting properties (midpoint of superconducting transition is $91.8$ K) for deposition of the YBCO films. As a reference, we deposited series of YBCO films on SAT-CAT-LA [(SrAl$_{0.5}$Ta$_{0.5}$O$_3$)$_{0.7}$(CaAl$_{0.5}$Ta$_{0.5}$O$_3$)$_{0.1}$(LaAlO$_3$)$_{0.2}$] substrates using the same PLD parameters. These substrates have in-plane lattice parameters well matched to the in-plane lattice parameters of YBCO and have been shown to produce exceptionally good quality films \cite{abal2}.

We have extended our procedure of thin YBCO film deposition developed for small size substrates to Al${\rm _2}$O${\rm _3}$ substrates of 2 inches diameter. To this end, we applied an excimer KrF pulsed laser and a vacuum system permitting not only target rotation but also target scanning and substrate rotation during film growth, in order to optimize both target erosion and film thickness uniformity \cite{PVD}. In this system the control of the substrate temperature is done using radiation heater, which is unlike the system with inconel heater block used for small-area films. Therefore, the appropriate deposition temperatures for both the buffer layers and YBCO films have been optimized anew. The deposition of CeO${\rm _2}$ layers, of about 30 nm thickness, was performed at a temperature of radiation heater of 820 $^\circ$C, at 300 mTr oxygen pressure, using a laser pulse energy of 160 mJ and a repetition rate of 5 Hz, applying 2500 laser pulses.  Next, the CeO${\rm _2}$ films were {\it ex situ} annealed in air at a temperature of 1000 $^\circ$C, for 12 hours, in a muffle furnace, at ambient pressure. Subsequently, the deposition of YBCO was performed at a temperature of radiation heater of 980 $^\circ$C, using a laser pulse energy of 200 mJ with repetition rate of 10 Hz, in flowing oxygen at a pressure of 300 mTr. Application of 13050 laser pulses resulted in the YBCO film thickness of about 200 nm. The post annealing process was much more elaborated than in the case of small substrate films, involving stepwise introducing of oxygen gas into the ablation chamber while stopping the temperature decrease at several temperatures. The large-area YBCO films were characterized by Cu-K$_\alpha$ X-ray diffraction.

The structural properties of the as-grown films were studied using Philips XPert Pro Alpha-1 MPD diffractometer and the surface topology was investigated by Digital Instruments atomic force microscope (AFM) MultiMode Nanoscope IIIA. The four-probe DC transport measurements were performed in the temperature range of 4.2 to 300 K, on YBCO films patterned by photolithography into 30 and 50 $\mu$m strips. We defined $T_{\rm{c0}}$ as the temperature at which $R/R_{\rm{N}}=0.1$, where $R$ is the film resistance and $R_{\rm{N}}$ is the normal-state resistance just above the onset of the superconducting transition, while the $j_{\rm{c}}$ was defined by the criterion of voltage drop exceeding $\sim$ $1$ $\mu$V at over 2 mm between the potential leads.

In addition, we have evaluated the critical current density of the films by two non-contact methods detecting local magnetic induction. First of these methods uses principle of the Faraday rotation in a magnetic sensor (gadolinium gallium garnet) placed directly on the top of the investigated sample in order to visualize the magnetic field penetration into superconducting film. Sample is placed inside a continuous-flow cryostat (temperature in the range of 4 - 300 K) equipped with a low-magnetic field magnet. The iron-garnet indicator rotates the plane of the polarized light proportionally to local magnetic field induction. The magneooptical image produced by the magnetic flux penetrating the film is subsequently used to calculate $j_{\rm{c}}$ by solving the inverse Biot-Savart problem \cite{Jooss}. The second method uses the Hall sensor placed in the film center, which detects the dependence of local magnetic induction on the film orientation with respect to the external magnetic field ($H = 2.5$ kOe). The sensor, of the area $20 \times 20$  $\mu$m$^2$, is a 2-dimensional electron gas device fabricated in a pseudomorphic AlGaAs/InGaAs/GaAs heterostructure. The local induction is used for the estimate of the $j_{\rm{c}}$ in the $T$-range from 4 to 80 K \cite{Gilchrist,Beek}.

\section{Finding optimal parameters of buffer and superconducting layers}

\subsection{CeO$_2$ buffer films: optimal recrystallization temperature}

\begin{figure}
\begin{center}
\includegraphics[width=8cm]{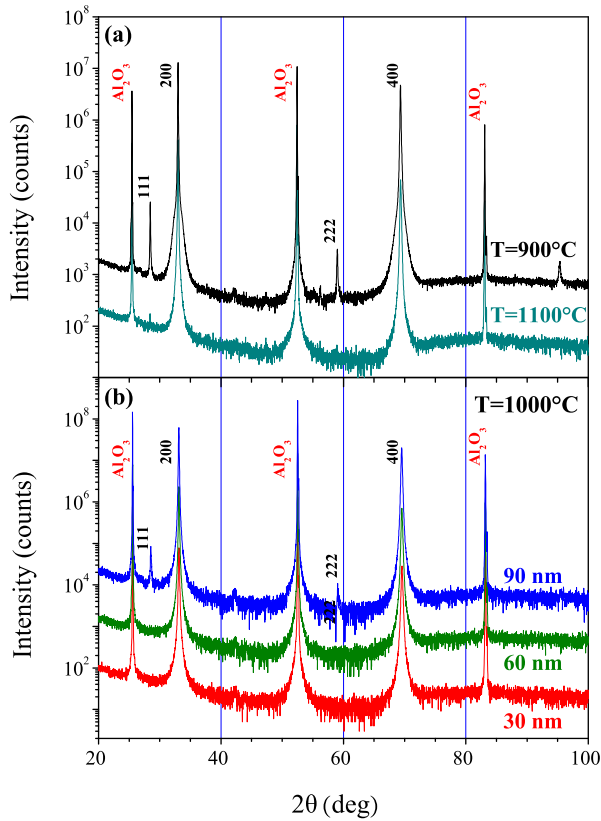}
\end{center}
\caption{(a) Comparison of structural properties of 200 nm CeO$_2$ films annealed at 900 and 1100 $^\circ$C. (b) The $\theta$-$2\theta$ scans of 30 nm, 60 nm and 90 nm CeO$_2$ films annealed at 1000 $^\circ$C.}
\label{Roentgen}
\end{figure}

The influence of the recrystallization temperature $T_{\rm{rec}}$ on the structural properties of CeO$_2$ films deposited on Al$_2$O$_3$ substrates has been evaluated in the first step of this study. For this purpose several CeO$_2$ films of the same thickness of 200 nm have been annealed at various $T_{\rm{rec}}$ in the range between 800 $^\circ$C and 1250 $^\circ$C. Figure \ref{Roentgen}(a) shows the $\theta$-$2\theta$ scans of two films annealed at 900 $^\circ$C and 1100 $^\circ$C. We observe that both scans show strong $l$ 00 diffraction peaks, indicating that majority of grains are c-axis aligned with c-axis perpendicular to the substrate plane. The film annealed at 900 $^\circ$C shows, in addition, small 111 and 222 peaks, resulting from small amount of misoriented grains. These peaks disappear when $T_{\rm{rec}}$ is increased above 900 $^\circ$C, so that the films become more homogeneous. 

\begin{figure}
\begin{center}
\includegraphics[width=8cm]{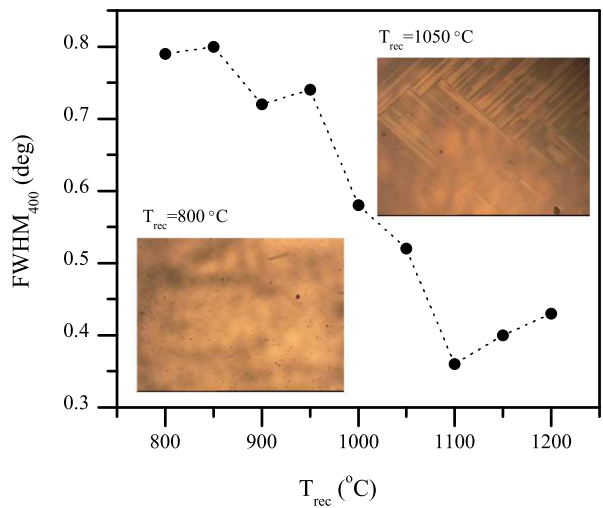}
\end{center}
\caption{The FWHM of the 400 diffraction peak for  200 nm thick CeO$_2$ films versus recrystallization temperature. Insets show surface photos from CCD camera of CeO$_2$ films annealed at 800 $^\circ$C (smooth film) and 1050 $^\circ$C (mosaic structure).}
\label{FWHM}
\end{figure}

Figure~\ref{FWHM} shows the full width at half maximum (FWHM) for 400 reflection peak, extracted from the $\theta$-$2\theta$ scans. We can see that the FWHM noticeably decreases with increasing $T_{\rm{rec}}$, which indicates that the structural order improves. With the help of AFM we have observed that this improvement in structural quality is accompanied by the appearance of pores on the CeO$_2$ film surface. Both the density and the size of pores increase with the increase of $T_{\rm{rec}}$, until cracks in the form of mosaic structure appear in films above $T_{\rm{rec}}$=1050 $^\circ$C. This mosaic structure is best visualized by a photos from CCD camera, as shown in two insets to figure~\ref{FWHM}. To avoid cracks, we have chosen the temperature $T_{\rm{rec}}$=1000 $^\circ$C as the optimal temperature for CeO$_2$ recrystallization.

\subsection{CeO$_2$ buffer films: optimal thickness}

In the next step we evaluate how the thickness of cerium dioxide layer affects the surface roughness of the buffer, since this parameter is crucial for a growth of good quality YBCO film. Figure \ref{Roentgen}(b) displays diffraction spectra of several CeO$_2$ films of different thicknesses, all annealed at the same temperature $T_{\rm{rec}}$ = 1000 $^\circ$C. We observe that films thicker than 60 nm exhibit additional 111 and 222 peaks, indicating the appearance of misoriented grains.

\begin{figure}
\begin{center}
\includegraphics[width=8cm]{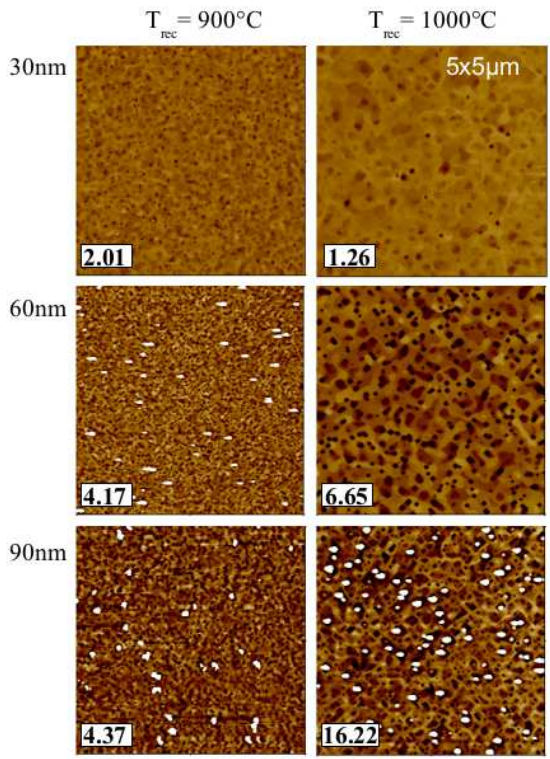}
\end{center}
\caption{AFM images of CeO$_2$ films of different thicknesses annealed at 900 $^\circ$C and 1000 $^\circ$C. The numbers in the lower left corners indicate root mean square roughness of these films in nm.}
\label{AFM}
\end{figure}

Figure~\ref{AFM} shows AFM images for films with thickness of 30 nm (upper row), 60 nm (middle row) and 90 nm (lower row), recrystallized at two different temperatures. The root mean square roughness ($\sigma_{\rm{RMS}}$) for the film area $5 \times 5$ $\mu$m$^2$ is indicated in the lower left corner of each image. Two top images show that in case of 30 nm thick films the $\sigma_{\rm{RMS}}$ value decreases as the recrystallization temperature increases, reaching the lowest value of 1.26 nm for a film recrystallized at $T_{\rm{rec}}$=1000 $^\circ$C. We also observe that further increase in buffer layer thickness and subsequent recrystallization does not result in a significant surface improvement, leading instead to an increase of both the porosity and $\sigma_{\rm{RMS}}$ values. Therefore, we conclude that the most promising is the use of 30 nm buffer layer, recrystallized at $T_{\rm{rec}}$=1000 $^\circ$C.

\begin{figure}
\begin{center}
\includegraphics[width=7cm]{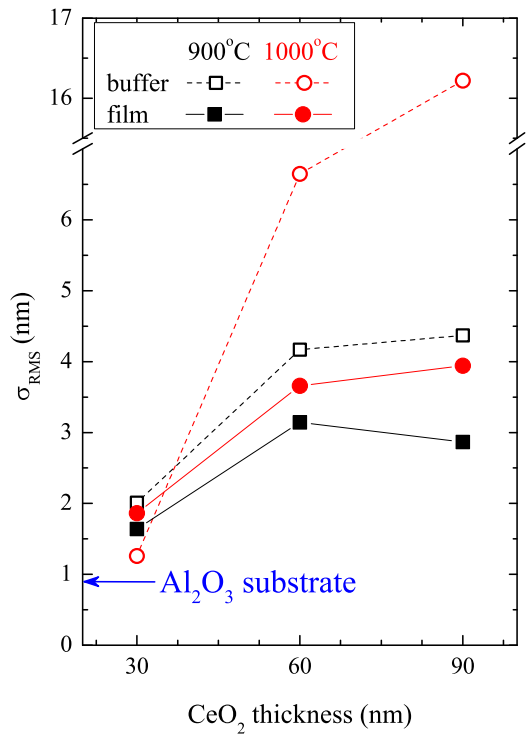}
\end{center}
\caption{Root mean square roughness versus buffer layer thickness for buffers (open points) annealed at 900 $^\circ$C (squares) and 1000 $^\circ$C (circles), and for 150 nm YBCO films (full points) subsequently deposited on the top of these buffer layers.}
\label{Roughness}
\end{figure}

Further, to verify the above conclusion, we have prepared the first series of superconducting films consisting of YBCO films deposited on the CeO$_2$-buffered sapphire. This series contains YBCO films of identical thickness of about 150 nm, deposited on the top of 30, 60 and 90 nm cerium dioxide layers. Figure~\ref{Roughness} summarizes $\sigma_{\rm{RMS}}$ for buffer layers and YBCO films grown on them. It demonstrates that the smoothest YBCO films grow on the thinnest (30 nm) CeO$_2$ buffer, as expected from the $\sigma_{\rm{RMS}}$ values of the buffer layers.

\begin{figure}\begin{center}
\includegraphics[width=7cm]{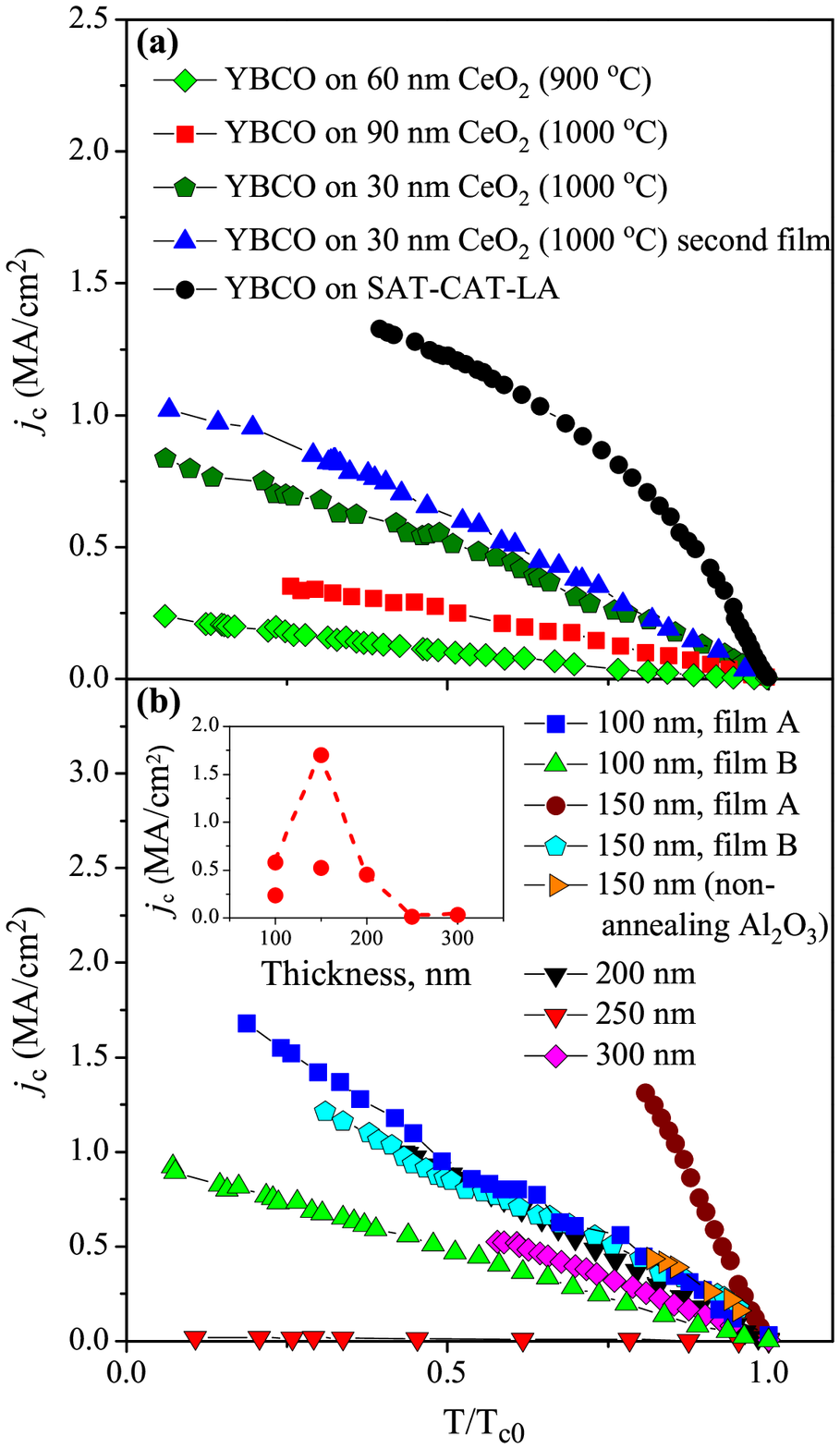}
\end{center}
\caption{(a) $j_{\rm{c}}$ versus $T/T_{\rm{c0}}$ for 150 nm YBCO films deposited on the top of 30, 60 and 90 nm CeO$_2$ buffer annealed at 900 $^\circ$C and 1000 $^\circ$C. (b) $j_{\rm{c}}$ versus $T/T_{\rm{c0}}$ for YBCO films of different thicknesses deposited on the top of 30 nm CeO$_2$ buffer annealed at 1000 $^\circ$C. Inset shows the $j_{\rm{c}}$ at $T/T_{\rm{c0}} = 0.75$ versus YBCO film thickness. Dashed line is drawn through the points with the highest $j_{\rm{c}}$ for each film thickness.}
\label{jc}
\end{figure}

Figure~\ref{jc}(a) shows the temperature dependence of transport $j_{\rm{c}}$ for some of these YBCO films, together with $j_{\rm{c}} (T)$ for YBCO films deposited directly on well-matched SAT-CAT-LA substrate. We observe that the $j_{\rm{c}}$ for films on CeO$_2$-buffered sapphire is much smaller compared to critical current density of YBCO films grown on well-matched substrate, in some cases by order of magnitude. However, the best values of $j_{\rm{c}}$ are obtained for films deposited on 30 nm buffer layers (recrystallized at $T_{\rm{rec}}$=1000 $^\circ$C), which confirms the conclusion that 30 nm is the best choice of thickness for a buffer layer.

\subsection{Optimal thickness of YBCO films}

In the following, in an attempt to optimize the thickness of YBCO films, we have deposited eight YBCO films of various thicknesses, ranging from 100 nm to 300 nm, on a 30 nm buffer layer, annealed at 1000 $^\circ$C.

Surprisingly, as we show in figure~\ref{jc}(b), the magnitude of the critical current density does not change monotonously with the increasing film thickness, but it is considerably scattered. In the inset we show the $j_{\rm{c}}$ at $T/T_{\rm{c0}}$=0.75, versus YBCO film thickness, and the dashed line is drawn through points with highest $j_{\rm{c}}$ for each thickness. We observe that the highest $j_{\rm{c}}$ is reached for thickness of 150 nm, and, as the films become thicker, the highest $j_{\rm{c}}$ decreases. This is consistent with the results observed earlier, attributed to the microcracking that occurs in YBCO films on CeO$_2$-buffered Al$_2$O$_3$ due to thermal strain during cooling, when the YBCO film thickness exceeds a value of about 250 - 300 nm \cite{microcracking}. We note also that the highest $j_{\rm{c}}$, which is observed for 150 nm thick film A, is larger than $j_{\rm{c}}$ in YBCO films grown on well-matched SAT-CAT-LA substrates. However, this large $j_{\rm{c}}$ is not reproduced in other 150 nm films; in addition, the scatter of the $j_{\rm{c}}$ values exists for other thicknesses. This indicates that there are other factors which affect film quality. One possibility is the different film oxygenation. In the following section we examine film reproducibility in more detail.

\section{Reproducibility and properties of YBCO films}

\begin{figure}
\begin{center}
\includegraphics[width=8cm]{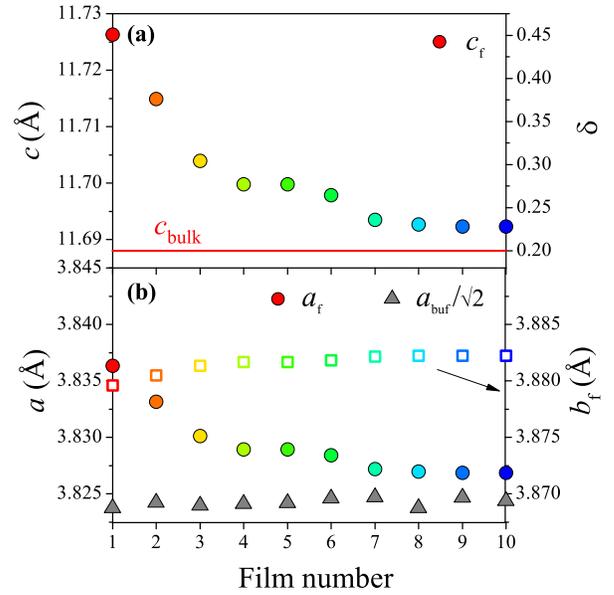}
\end{center}
\caption{(a) Lattice constant $c_{\rm{f}}$ for set of ten YBCO films, 150 nm thick, on CeO$_2$ buffered sapphire. The red line indicates $c$ parameter for bulk YBCO with $\delta = 0.2$. The right scale shows the oxygen deficiency $\delta$, determined using relation from reference \cite{Kuru}. (b) The buffer lattice parameter $a_{\rm{buf}}/\sqrt{2}$ (grey triangles, left scale), measured for all films, and in-plane YBCO parameters, color-coded as in (a), $a_{\rm{f}}$ (circles, left scale) and $b_{\rm{f}}$ (squares, right scale), calculated as described in the text.}
\label{c}
\end{figure}

In order to investigate film reproducilibity, we have measured structural, microstructural and transport properties of set of ten YBCO films, 150 nm thick, grown under identical conditions. All films show X-ray diffraction patterns of YBCO of good-quality, with a $c$-axis perpendicular to the substrate plane, and negligibly small volume fraction of misaligned grains. However, the films differ slightly in the value of the $c$-axis lattice parameter, $c_{\rm{f}}$. We observe that the deviation of the $c_{\rm{f}}$ from the bulk value is quite random, indicating that the variation is not caused by a deterioration of the deposition parameters during the growth of ten consecutive samples. To simplify further discussion, we enumerate the films in the order of decreasing $c_{\rm{f}}$ parameter, determined from the 007 X-ray diffraction peak, as shown in figure~\ref{c}(a) (we stress that this is not the order, in which the films were deposited). We observe that in all films the $c_{\rm{f}}$ is slightly larger than the $c$ parameter in the bulk YBa$_2$Cu$_3$O$_{7-\delta}$ with $\delta=0.2$, which is indicated by red line in the figure.

\begin{figure}\begin{center}
\includegraphics[width=7.5cm]{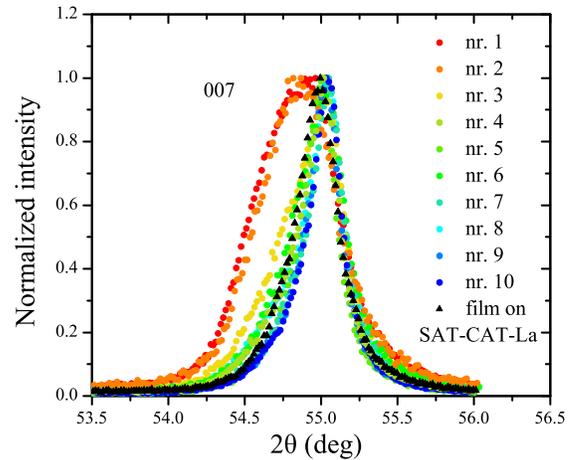}
\end{center}
\caption{The enlarged view of $\theta$-$2\theta$ scans in the vicinity of 007 peak for a set of YBCO films (the data are color-coded as in figure~\ref{c}(a)).}
\label{007}
\end{figure}

The variable expansion of the $c_{\rm{f}}$ may originate from a combination of two effects. The first effect is a small compressive in-plane strain in the YBCO lattice cell (which grows rotated by 45$^{\circ}$ in the CeO$_2$ basal plane), caused by a mismatch between lattice constant of CeO$_2$ buffer layer, $a_{\rm{buf}}$, and the in-plane lattice constants of YBCO (which are equal to $a = 3.826$ {\AA} and $b = 3.883$ {\AA}, respectively, for YBCO with $\delta=0.2$). The $a_{\rm{buf}}$ divided by $\sqrt{2}$ (for easier comparison with YBCO parameters), measured for all buffer layers is shown in figure~\ref{c}(b) by grey triangles; it is smaller than the in-plane lattice parameters of YBCO. While such compressive strain may lead to some expansion of the $c_{\rm{f}}$, the variation of $a_{\rm{buf}}$ from film to film is smaller than 0.06\%, so it cannot be the main cause of large changes of the $c_{\rm{f}}$, which exceed 0.3\%.

The second effect is a variable oxygen deficiency $\delta$ in YBa$_2$Cu$_3$O$_{7-\delta}$, which leads to an increase of $c$ lattice parameters \cite{Gallagher}. This effect is well known and has been used in the past for nondestructive evaluation of $\delta$ in bulk crystals \cite{Benzi}. It has been observed in case of YBCO films (200 nm thick, deposited on SrTiO$_3$ substrates) that the dependencies of lattice parameters on the oxygen deficiency follow the linear dependencies observed in the bulk \cite{Kuru}, $c({\delta}) = 11.657 + 0.154\cdot{\delta} $, $a({\delta}) = 3.817+0.043\cdot{\delta} $, and $b({\delta}) = 3.885-0.012\cdot{\delta}$. Using these relations we deduce that in the present case $\delta$ varies between 0.23 and 0.45, as indicated on the right scale in figure~\ref{c}(a); furthermore, we may extract the values of in-plane parameters, $a_{\rm{f}}$ and $b_{\rm{f}}$, which are shown in figure~\ref{c}(b).

\begin{figure}\begin{center}
\includegraphics[width=8cm]{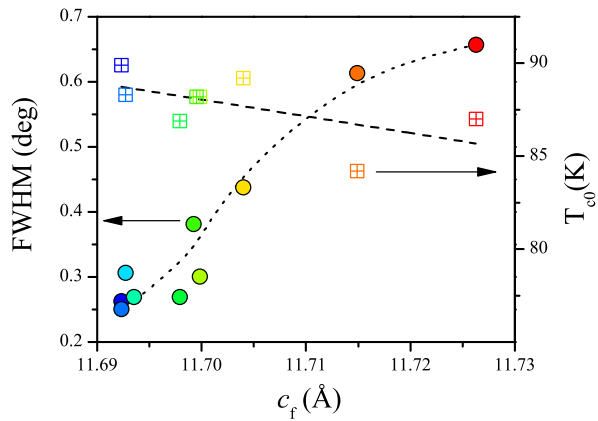}
\end{center}
\caption{The FWHM of 007 diffraction peak (left scale, full circles and dotted line) and $T_{\rm{c0}}$ (right scale, squares and dashed line) versus $c_{\rm{f}}$ for a set of YBCO films. The data are color-coded as in figure~\ref{c}(a). The lines are guides to the eye.}
\label{FWHM2}
\end{figure}

The oxygen deficiency in the YBCO films is further confirmed by the shape of the 007 X-ray diffraction peak, displayed in figure~\ref{007}, and the correlation between the FWHM for this peak and the $c_{\rm{f}}$ parameter, shown in figure~\ref{FWHM2} (left scale, full circles). The peak is narrow but asymmetric in all samples, in which $c_{\rm{f}}$ is close to the bulk value, with broad "tail" which extends towards lower values of 2$\theta$. Such asymmetry suggests that in the initial stage of the film deposition large oxygen deficiency exists, so that $c_{\rm{f}}$ is large. It is possible that the small compressive strain contributes somewhat to the expansion of the $c_{\rm{f}}$ in this initial stage. However, with increasing film thickness films become better oxygenated, the strain is relaxed, and $c_{\rm{f}}$ decreases. On the other hand, the 007 peak is wider but symmetrical in the samples with the largest $c_{\rm{f}}$-value (films '1' and '2'). Thus, these films appear to be uniformly oxygen-deficient throughout the film thickness.

\begin{figure}
\begin{center}
\includegraphics[width=8cm]{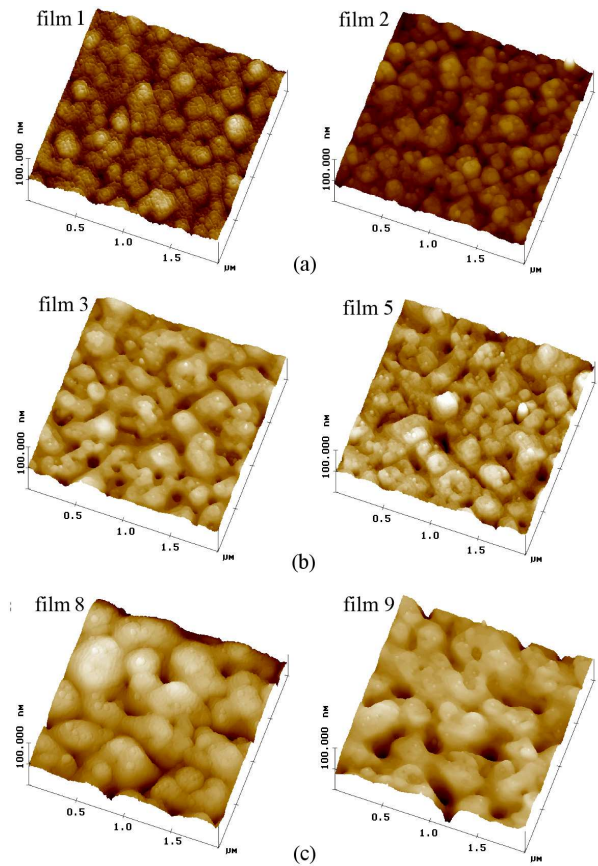}
\end{center}
\caption{AFM images ($2 \times 2$ $\mu$m$^2$ area) of YBCO films. (a) Small (10-100 nm) grains. (b) Medium (100-200 nm) grains. (c) Large (200-500 nm) grains.}
\label{AFM2}
\end{figure}

Figure~\ref{AFM2} shows AFM images for several representative films. According to the images, the film growth begins by the nucleation of islands, which then merge to form solid films. Island nucleation is most likely initiated by structural defects in the buffer layer. During recrystallization of CeO$_2$ coalescing grains form a film with a low value of $\sigma_{\rm{RMS}}$, about 1.3 nm, but with a set of shallow cavities on the buffer layer surface. In addition, the images show that the grain size in YBCO films varies from small (10-100 nm) in films '1' and '2', through intermediate (100-200 nm) in films '3' and '5', to large (200-500 nm) in films '8' and '9'. The diverse grain sizes may originate in different surface roughness of the buffer layers, which is likely to vary slightly form film to film in spite of nominally the same conditions of the buffer growth and recrystallization. Increased buffer surface roughness may lead to increased density of island nuclei, which in turn will lead to smaller grain size.

It is noteworthy that there is a distinct correlation between the size of grains in YBCO films and the value of $c_{\rm{f}}$. In particular, films '1' and '2' with the largest $c_{\rm{f}}$ values consist of the smallest grains. On the other hand, the films '8' and '9' with largest grains display $c_{\rm{f}}$ closest to the bulk value and accompanied by relaxation of the $c_{\rm{f}}$ across the film thickness. It is possible that large voids at the grain boundaries, which exist in the films with large grains, contribute to better oxygenation during film growth and therefore lead to a decrease in $c_{\rm{f}}$.

We now discuss the superconducting properties of YBCO films. Figure~\ref{FWHM2} (right scale, squares) shows the dependence of $T_{\rm{c0}}$ on the $c_{\rm{f}}$ parameter. The dashed line shows that on average $T_{\rm{c0}}$ decreases as $c_{\rm{f}}$ increases, as might be expected from the increasing oxygen deficiency. However, there is a considerable spread of data around the dashed line, which indicates that in addition to oxygen deficiency, there is some other factor affecting the value of $T_{\rm{c0}}$. It is very likely that this factor may be the grain boundary scattering, which should be larger when the grains are small and the grain boundary density is high.

\begin{figure}\begin{center}
\includegraphics[width=7cm]{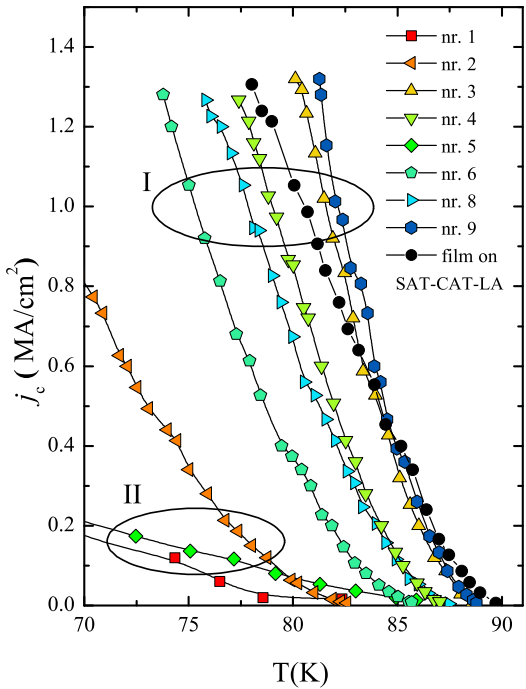}
\end{center}
\caption{The $T$-dependence of $j_{\rm{c}}$ for several YBCO films. 'I' and 'II' denotes two groups of films with high and low $j_{\rm{c}}$, respectively. The data points are color-coded as in figure~\ref{c}(a).}
\label{j_c}
\end{figure}

Figure~\ref{j_c} shows the dependence of $j_{\rm{c}}$ on temperature, measured by transport method for most of the ten films deposited on CeO$_2$ and for one film grown on a SAT-CAT-LA substrate. All films may be divided into two groups. The group I contains films in which $j_{\rm{c}}$ grows quite rapidly with decreasing $T$. All films in this group exhibit small $c_{\rm{f}}$ parameter (the sample with the steepest increase of $j_{\rm{c}}$ is the film '9' with the smallest $c_{\rm{f}}$).The group II consists of films with a much smaller value of the $j_{\rm{c}}$, and two of these films have a high value of the $c_{\rm{f}}$. Therefore, it seems that the $j_{\rm{c}}$ follows similar trend as $T_{\rm{c0}}$, that is, the $j_{\rm{c}}$ decreases as $c_{\rm{f}}$ grows. This indicates that the oxygen deficiency contributes to the suppression of the critical current density. However, just as it is in the case of the correlation between $T_{\rm{c0}}$ and the $c_{\rm{f}}$, we observe no perfect one-to-one correlation between the $j_{\rm{c}}$ and the $c_{\rm{f}}$. For example, while $c_{\rm{f}}$ (and oxygen deficiency) is larger for film '3' than for film '5', the $j_{\rm{c}}$ in film '5' is much smaller than in film '3'. This leads to the conclusion that the film microstructure, besides the oxygen deficiency, must be responsible for the magnitude of the $j_{\rm{c}}$. We recall in this context that greater surface roughness of CeO$_2$ buffer has been reported to depress the $j_{\rm{c}}$ \cite{Wang}. This would be consistent with our observation that $j_{\rm{c}}$ correlates to some extent with the grain size, which most likely is determined by the density on defects on the buffer surface. It is important to remember, however, that the variation of the density of defects on the buffer surface may affect not only the density of grain boundaries (which we observe), but may also lead to variable density of other defects which influence the $j_{\rm{c}}$, such as, for example, stacking defects or second phase precipitates. We will comment at the end of this section on the possible type of defects which affect the $j_{\rm{c}}$ in the present case.

\begin{figure}\begin{center}
\includegraphics[width=7cm]{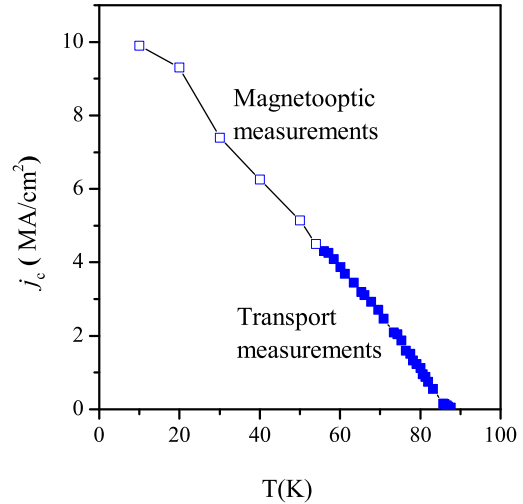}
\end{center}
\caption{The $T$-dependence of $j_{\rm{c}}$ for YBCO film measured by magneto-optical (open symbols) and transport (full symbols) methods.}
\label{MO}
\end{figure}

Transport measurements of the $j_{\rm{c}}$ were performed at high temperatures just below the $T_c$ due to the limitation of the maximum current density used in the measuring setup. In order to evaluate low-temperature behavior we use magneto-optical method of visualization of magnetic flux penetration into the film. The optical images of flux are subsequently converted  into distribution of critical current density in the sample using the Biot-Savart inversion procedure \cite{Jooss}. $j_{\rm{c}}$ was defined as a plateau value in the resulting critical current density distribution. Magneto-optical measurements show that the $j_{\rm{c}}$ increases linearly with decreasing $T$ down to 4 K, as shown in figure~\ref{MO}. The linear dependence of the $j_{\rm{c}}$ on temperature is predicted in case of strong pinning by correlated defects. This is in contrast to weak collective pinning, which would manifest itself as an exponential decay of $j_{\rm{c}}$ with an increase of $T$ \cite{Beek2,Polat}.

The YBCO for device applications should be characterized by high values of the $j_{\rm{c}}$ at temperatures of $T \simeq 77$ K. By extrapolating the data from figure~\ref{j_c} to a temperature of $T = 77$ K we obtain $j_{\rm{c}}$ values in the range of 1.35 to 2.4 MAcm$^{-2}$ for the three best films deposited on CeO$_2$, while $j_{\rm{c}}$ for the film deposited on SAT-CAT-LA substrate is about 1.5 MAcm$^{-2}$. Thus, the value of $j_{\rm{c}}$ for the best YBCO films grown on CeO$_2$ is greater than that obtained for a well matched substrate. Such a value of $j_{\rm{c}}$ is suitable for the applications of these films in microwave devices and fault current limiters.

\begin{figure}\begin{center}
\includegraphics[width=8cm]{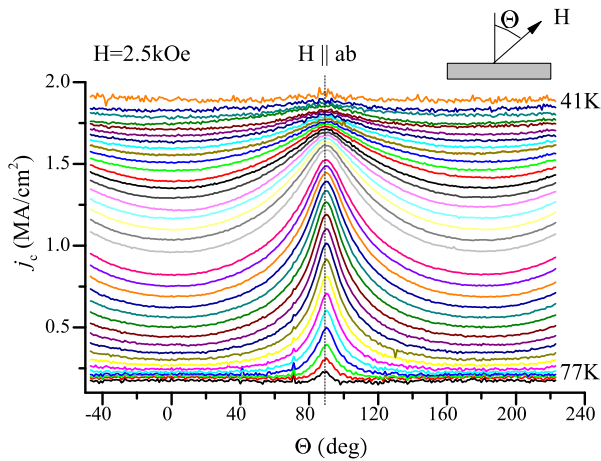}
\end{center}
\caption{The dependence of $j_{\rm{c}}$ on the field orientation angle $\Theta$, measured for temperatures between 41 and 77 K, at intervals of 2 K. The $\Theta$ is the angle between the field direction and perpendicular to the sample plane.}
\label{angle}
\end{figure}

In order to evaluate the possible origin of strong pinning in our films, it is useful to study the dependence of the $j_{\rm{c}}$ on the orientation of the external magnetic field $H$, as shown in figure~\ref{angle} ($\Theta$ is the angle between the direction of $H$, and perpendicular to the film plane). It is evident that there is only one sharp peak on the $j_{\rm{c}}$($\Theta$) curves, situated at $\Theta$ = $90^{\circ}$, that is, for the magnetic field rotated parallel to the ab plane ($H || ab$). Such peaks, observed for $H || ab$, have been reported in case of most of thin YBCO films, grown on various substrates  \cite{Polat,Kees}. They indicate strong pinning by correlated disorder caused by a layering near the $ab$ plane of second phase precipitations, intergrowths or stacking defects. We note that in one of the previous studies of YBCO films grown on CeO$_2$, in addition to the peaks for $H || ab$, an enhancement of $j_{\rm{c}}$ has been observed for magnetic field parallel to the $c$ axis. This has been attributed to the self-assembled nano-dots on the CeO$_2$ buffer surface \cite{Nie2}. In the present study we do not observe such enhancement. This does not mean that the defects related to the buffer layer do not play any role, but only that they do not contribute to correlated disorder away from the $ab$ plane. Indeed, the correlation between the $j_{\rm{c}}$ and the film microstructure suggests that the structural defects induced by a buffer layer, together with the oxygen content, determine the magnitude of the $j_{\rm{c}}$.

\section{Microwave filters on large area substrates}

\begin{figure}
\centering
\includegraphics[width=7cm]{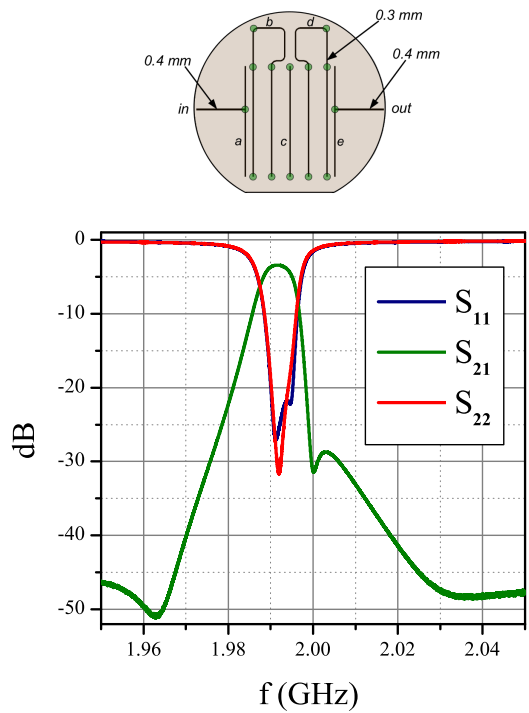}
\caption{Top part: filter geometry, with individual half-wave resonators labeled by {\it a}, {\it b}, {\it c}, {\it d}, and {\it e}. Bottom: the transmission coefficient S$_{21}$ and both reflections coefficients, S$_{11}$ and S$_{22}$, respectively, of the narrow-band microwave filter, measured at liquid nitrogen temperature.} \label{Filter}
\end{figure}

In order to verify the suitability of our films for microwave applications, we have fabricated a narrow-band microstrip filter using a 2 inch diameter YBCO film deposited on one side of a CeO$_2$ buffered r-cut oriented sapphire substrate. A thin silver or gold film ground plane has been evaporated on the opposite side of the substrate, using standard resistively heated source of a vacuum evaporator, after the thermal treatment of both CeO$_2$ and YBCO has been completed. The filter geometry was defined using photolithography, followed by wet etching of the YBCO film by diluted citric acid.

A sketch of the filter geometry is shown in the top part of Figure \ref{Filter}. The filter is built of 5 coupled microstrip half-wave resonators, with input and output lines connected to resonators {\it a} and {\it e}, respectively. Resonators {\it a}, {\it c} and {\it e} have the same length and are parallel to each other, whereas resonators {\it b} and {\it d} are three times longer. The red dots denote the positions of 14 dielectric tuners. The filter is mounted inside of a copper housing.

The performance of this filter, characterized in a liquid nitrogen bath using a microwave network analyzer with attached semirigid stainless steel cables, was already reported \cite{Abramowicz}. In brief, the designed filter center frequency was 2000 MHz, its bandwidth 5 MHz, and the return loss better than -20 dB. A typical filter performance, after tuning, is shown in bottom part of Figure \ref{Filter}. The measured transmission  coefficient S$_{21}$ and both reflection coefficients, S$_{11}$ and S$_{22}$, respectively, of the filter are in good agreement with the results of full wave electromagnetic simulations, performed using a commercially available software package \cite{QuickWave}.

\section{Conclusions}

We investigated the properties of CeO$_2$ films grown on r-cut sapphire substrates in order to obtain buffer layers suitable for the deposition of YBCO films characterized by very good superconducting parameters. 30 nm buffer layers recrystallized at $1000 ^\circ$C have been found to have the best structural properties.  We deposited a set of YBCO films on top of 30 nm CeO$_2$ buffer layers and found a significant correlation between the film microstructural properties and the oxygen content in the films and the superconducting parameters, $T_{\rm{c0}}$, and $j_{\rm{c}}$. The best films with the highest $j_{\rm{c}}$ value suitable for application were those in which the strain introduced by the buffer layer relaxes and large grains are formed during growth.

\section*{Acknowledgements}

This work was supported by the European Union within the European Regional Development Fund, through the Innovative Economy grant POIG.01.01.02-00-108/09 and by Project APVV-0494-11, the project CENTE II, R\&D Operational Program funded by the ERDF, ITMS code 26240120019.


\begin{thebibliography}{99}

\bibitem{Yamasaki} H. Yamasaki, M. Furuse, Y. Nakagawa, Appl. Phys. Lett. {\bf85}, 4427 (2004).
\bibitem{Zhao} H. Zhao, X. Wang, J. Z. Wu, Supercond. Sci. Technol. {\b21}, 085012 (2008).
\bibitem{SIMON} R. W. Simon, R. B. Hammond, S. J. Berkowitz, B. A. Willemsen, Proceedings of the IEEE {\bf92}, 1585 (2004).
\bibitem{Lorenz} M. Lorenz, H. Hochmuth, D. Natusch, M. Kusunoki, V. L. Svetchnikov, V. Riede, I. Stanca, G. K\"{a}stner, D. Hesse, IEEE Trans. Appl. Supercond. {\bf11}, 3209 (2001).
\bibitem{Wu} X. D. Wu, R. C. Dye, R. E. Muenchausen, S. R. Foltyn, M. Maley, A. D. Rollett, A. R. Garcia, and N. S. Nogar, Appl. Phys. Lett. {\bf58}, 2165 (1991).
\bibitem{Chromik1} M. Spankova, I. Vavra, S. Gazi, D. Machajdik, S. Chromik,  K. Frohlich, L. Hellemans, S. Benacka, Journal of Crystal Growth {\bf218}, 287 (2000).
\bibitem{Ohki} K. Ohki, K. Develos-Bagarinao, H. Yamasaki, Y. Nakagawa, Journal of Physics: Conference Series {\bf 97}, 012142 (2008).
\bibitem{Nie} J. C. Nie, H. Yamasaki, Y. Nakagawa, K. Develos-Bagarinao, M. Murugesan, H. Obara, Y. Mawatari, Journal of Physics: Conference Series {\bf43}, 353 (2006).
\bibitem{microcracking} A.G. Zaitsev, G. Ockenfuss, R. Wördenweber, Inst. Phys. Conf. Ser., Vol. 158, 3rd European Conference on Applied Superconductivity, IOP Publishing Ltd 1997, p. 25.
\bibitem{Kastner} G. Kastner, D. Hesse, M. Lorenz, R. Scholz, N. D. Zakharov, P. Kopperschmidt, Phys. Stat. Sol. (a) {\bf150}, 381 (1995).
\bibitem{Develos-Bagarinao} K. Develos-Bagarinao, H. Yamasaki, Y. Nakagawa, H. Obara, H.Yamada, Physica C {\bf 392–396}, 1229 (2003).
\bibitem{abali} I. Abal'osheva, I. Zaytseva, M. Aleszkiewicz, Y. Syryanyy, P.Gier\l{}owski, O. Abal'oshev, V. Bezusyy, M. Z. Cieplak, Acta Phys. Pol. A, {\bf 121}, 805 (2012)
\bibitem{abali1} I. Abal'osheva, I. Zaytseva, M. Aleszkiewicz, A. Malinowski, V. Bezusyy, Y. Syryanyy, P. Gier\l{}owski, O. Abal'oshev, M.Konczykowski, M. Z. Cieplak, {\bf 126}, A-69 (2014).
\bibitem{abal2} I. S. Abal'osheva, M. Z. Cieplak, Z. Adamus, M. Berkowski, V. Domukhovski, M. Aleszkiewicz, Acta Phys. Pol. A, {\bf 109}, 549 (2006).
\bibitem{PVD}PLD/MBE-2000 Deposition System, PVD Products, Inc (USA).
\bibitem {Jooss} Ch. Jooss, A. Forkl, R. Warthmann, H. Kronm\"{u}ller, Physica C {\bf 299}, 215 (1998).
\bibitem{Gilchrist} J. Gilchrist, M. Konczykowski, Physica C {\bf 212}, 43 (1993).
\bibitem{Beek} C. J. van der Beek, M. Konczykowski, V. M. Vinokur, G. W. Crabtree, T. W. Li, and P. H. Kes, Phys. Rev. B {\bf 51}, 15492 (1995).
\bibitem{Gallagher} P. K. Gallagher, H. M. O'Bryan, S. A. Sunshine, D. W. Murphy, Mat. Res. Bull. {\bf 22}, 995 (1987).
\bibitem{Benzi} P. Benzi, E. Bottizzo, N. Rizzi, J.Cryst.Growth {\bf 269}, 625 (2004).
\bibitem{Kuru} Y. Kuru, M. Usman, G. Cristiani, H.-U. Habermeier, J. Crystal Growth {\bf 312}, 2904 (2010).
\bibitem{Wang} B. Wang, L. Liu, X. Wu, Y. Yao, M. Wang, S. Lu, and Y. Li, J. Supercond. Nov. Magn. {\bf 29}, 2487 (2016).
\bibitem{Beek2} C. J. van der Beek, M. Konczykowski, V. M. Vinokur, T. W. Li, P. H. Kes, and G. W. Crabtree, Phys. Rev. Lett. {\bf 74}, 1214 (1995).
\bibitem{Polat} \"{O}. Polat, J. W. Sinclair, Y. L. Zuev, J. R. Thompson, D. K. Christen, S. W. Cook, D. Kumar, Y. Chen, and V. Selvamanickam, Phys. Rev. B {\bf 84}, 024519 (2011).
\bibitem{Kees} C. J. van der Beek, M. Konczykowski, A. Abal'oshev, I. Abal'osheva, P. Gier{\l}owski, S. J. Lewandowski, M. V. Indenbom, and S. Barbanera,  Phys. Rev. B {\bf 66}, 024523 (2002).
\bibitem{Nie2} J. C. Nie, H. Yamasaki, H. Yamada, Y. Nakagawa, K. Develos-Bagarinao and Y. Mawatari, Supercond. Sci. Technol. {\bf 17}, 845 (2004).
\bibitem{Abramowicz}A Abramowicz, P Gierlowski, and M Jaworski, 2016 21st International Conference on Microwave, Radar and Wireless Communications (MIKON), DOI:10.1109/MIKON.2016.7492042
\bibitem{QuickWave}QuickWave v 7.5, www.qwed.eu

\end{thebibliography}
\end{document}